\documentclass[journal]{IEEEtran}
\usepackage{cite}

\ifCLASSINFOpdf
   \usepackage[pdftex]{graphicx}
   \graphicspath{{../pdf/}{../jpeg/}{./pdf}}
  \DeclareGraphicsExtensions{.pdf,.jpeg,.png}
\else
   \usepackage[dvips]{graphicx}
   \graphicspath{{../eps/}}
   \DeclareGraphicsExtensions{.eps}
\fi
\usepackage{amsmath}
\usepackage{ntheorem}
\theoremheaderfont{\bf}
\theorembodyfont{\rm}
\theoremseparator{:}

\usepackage{array}
\newcommand{\PreserveBackslash}[1]{\let\temp=\\#1\let\\=\temp}
\newcolumntype{C}[1]{>{\PreserveBackslash\centering}p{#1}}
\newcolumntype{R}[1]{>{\PreserveBackslash\raggedleft}p{#1}}
\newcolumntype{L}[1]{>{\PreserveBackslash\raggedright}p{#1}}

\usepackage{subfigure}
\ifCLASSOPTIONcompsoc
  \usepackage[caption=false,font=normalsize,labelfont=sf,textfont=sf]{subfig}
\else
  \usepackage[caption=false,font=footnotesize]{subfig}
\fi
\usepackage{url}

\usepackage{multirow}
\usepackage[justification=centering, font=small]{caption} 
\usepackage{bm}
\begin{document}
%
\title{A Method for Low Complexity Successive Cancellation List Decoding of Polar Codes}
%
%
%

\author{Liang~Ma,
        Hang~Li,
        and~Yuejun~Wei, \\
\thanks{L. Ma, H, Li and Y. Wei are with the Department of Wireless Research Institution, Shanghai Huawei Technologies Co., Ltd., Shanghai 201206, China (e-mail: maliang9@huawei.com; lihang20@huawei.com; weiyuejun@huawei.com).}

}

%
%

\markboth{}%
{Shell \MakeLowercase{\textit{et al.}}: Bare Demo of IEEEtran.cls for IEEE Journals}
%



\maketitle

\begin{abstract}
Polar codes have attracted a lot of attention during past few years and have been adopted as a coding scheme for 5G standard. Successive-cancellation list (SCL) decoder provides high level error-correction performance for polar codes, but the implementation complexity grows rapidly with the increase of the list size. Since the computation cost of sorting, many works focus on reducing the sorting complexity for SCL decoder. In this paper, we propose a method for SCL which directly reduce the number of input elements of sorting network without performance loss. Compared with SCL decoder, the proposed method has up to 95\% less sorter size on average, and the performance loss is negligible.
\end{abstract}

\begin{IEEEkeywords}
Polar codes, successive cancellations list decoding, sorter.
\end{IEEEkeywords}

%
\IEEEpeerreviewmaketitle

\section{Introduction}
%
%
%
%
\IEEEPARstart{P}{olar} codes \cite{bibitem01},  as the first class of capacity achievable codes, are applied as the coding scheme of the control channels in enhanced mobile broadband (eMBB) in the 5G wireless communication system \cite{bibitem11}. Successive cancellation (SC) algorithm is proposed in \cite{bibitem01} with low complexity, however, it has a big performance gap compared with ML bound. In \cite{bibitem02}, SC list (SCL) decoder is proposed which significantly narrows down the performance gap. However, a large list number is required for SCL algorithm in order to achieve a relative better performance, which increases the decoding latency and resource overhead for realization.

In order to decrease the decoding latency, an SC and ML hybrid decoder is proposed \cite{bibitem03, bibitem04}. But the complexity is extremely high for implementation. Another method is to increase the degree of decoding parallelism by introducing specific node types, in which multiple bits are decoded simultaneously. To be specific, some simplified path metric (PM) calculations of special nodes (Rate-1, Rate-0, SPC and Rep) are adopted in \cite{bibitem08}. For multi-bit parallel decoding, the number of splitting pathes grows exponentially with the number of information bits in the node, resulting in high computation complexity of path metric sorting. 

PM sorting is studied in many literatures. Bitonic sorting network, with easy implementation and short delay, can be adopted in SCL decoder to select paths. By taking advantage of the relation of the input value, \cite{bibitem05, bibitem06} proposed a simplified bitonic sorter in which redundant comparators are omitted. But the simplified bitonic sorter is difficult to be adopted for multi-bit parallel decoding. A radix sorting based state machine is proposed in \cite{bibitem07}, which has lower complexity for high channel quality. \cite{bibitem09, bibitem10} significantly reduce the sorting network size by limiting the number of splitting pathes in Rate-1 and SPC node. But this method introduces slight performance loss and cannot process the node other than Rate-1 and SPC. Therefore, further investigation is needed to reduce the complexity of sorting network.

This paper focus on reducing sorting network size of SCL decoder. Since the input LLRs for SCL nodes are partially ordered, some paths are more likely to be valid than others. Here, we propose a method for polar SCL decoder, which uses the partial ordered property to find the paths more likely to be correct. Besides, a pattern based selection method is also proposed to further decrease the computation complexity.

The rest of the paper is organized as follows. Section \uppercase\expandafter{\romannumeral2} provides a background on polar codes and its decoding algorithms. In Section \uppercase\expandafter{\romannumeral3}, we propose an simplified SCL polar decoder. Numerical results will be shown in section \uppercase\expandafter{\romannumeral4}. Finally, conclusions are drawn in Section \uppercase\expandafter{\romannumeral5}.

\section{Preliminaries}


\subsection{Polar Codes}
Polar code is a linear block code recursively concatenated by Kronecker power. An $(N,K)$ polar code can be represented as
\begin{equation}
{\bm{x}} = {\bm{u}}{\bm{G}^{ \otimes n}}
\label{equation01}
\end{equation}
where ${\bm{u}} = \{ {u_0},{u_1},...,{u_{N - 1}}\}$ is the input vector, ${\bm{x}} = \{ {x_0},{x_1},...,{x_{N - 1}}\}$ is the encoded vector, $\bm{G} = \left[ {\begin{array}{*{20}{c}}
1&0\\
1&1
\end{array}} \right]$ and $\otimes n$ denotes nth Kronecker power.

The input vector ${\bm{u}}$ is comprised of $K$ information bits and $N-K$ frozen bits. The frozen bits are usually set to be a predefined value (typically 0) known by the decoder. Vector ${\bm{x}}$ is transmitted through the channel, and the decoder receivers the Logarithmic Likelihood Ratio (LLR) vector ${\bm{y}} = \{ {y_0},{y_1},...,{y_{N - 1}}\}$ for decoding.

\subsection{Successive Cancellation (SC) and Successive Cancellation list (SCL) decoding}
SC decoding can be represented as a binary tree search, and each node represents a sub codeword \cite{bibitem08}. The LLRs which is represented as $\alpha$ go through the nodes from parents to children, and the hard estimates $\beta$ pass from child to parent. The left and right message ${\alpha ^l}$ and ${\alpha ^r}$ are calculated as \cite{bibitem08}
\begin{equation}
\begin{array}{l}
\alpha _i^l = \ln \left( {\frac{{1 + {e^{{\alpha _i} + {\alpha _{i + {2^{s - 1}}}}}}}}{{{e^{{\alpha _i}}} + {e^{{\alpha _{i + {2^{s - 1}}}}}}}}} \right)\\
\alpha _i^r = {\alpha _{i + {2^{s - 1}}}} + (1 - 2\beta _i^l){\alpha _i}
\end{array}
\label{equation02}
\end{equation}
While $\beta$ is computed as
\begin{equation}
{\beta _i} = \left\{ {
\begin{array}{ll}
\beta _i^l \oplus \beta _i^r,&if{\quad}i < {2^{s - 1}}\\
\beta _{i - {2^{s - 1}}}^r,&otherwise
\end{array}} \right.
\label{equation03}
\end{equation}
Where $s$ denotes the node stage in binary tree, $ \oplus $ denotes the bitwise XOR, and $i$ represent the index for current decoded bit in vector ${\bm{u}}$. Due to the data dependency, each node receives $\alpha$ first, than calculates ${\alpha ^l}$, hard decides ${\beta ^l}$, calculates ${\alpha ^r}$, hard decides ${\beta ^r}$, and finally sends ${\beta}$.

\cite{bibitem02} proposed SCL decoding algorithm to improve the error correction performance. The key idea is keep not only most reliable decoded sequence but also some suboptimal decoded sequences. To this purpose, a path metric (PM) is associated to each path and update at every new estimation as a cost function. For each decoding step, SCL stores a reliability metric $PM_l^i$ for each path $l$ that is updated for every estimated bit $i$ according to:
\begin{equation}
PM_l^i = \left\{
 \begin{array}{ll}
PM_l^{i - 1} + \left| {{\alpha _{{i_l}}}} \right|, & if\;{{\hat u}_{{i_l}}} \ne \frac{1}{2}(1 - {\mathop{\rm sgn}} ({\alpha _{{i_l}}})),\\
PM_l^{i - 1}, & otherwise,
\end{array} 
\right.
\label{equation04}
\end{equation}
Where $l$ is the path index, ${\alpha _{{i_l}}}$ is the calculated LLR value of bit $i$ at path $l$, and ${\hat u_{{i_l}}}$ is the estimate of bit $i$ at path $l$.

\subsection{Simplified SCL}
In [9], some special nodes with constituent bits are identified, and the candidate codewords corresponding to these nodes can be directly generated without traversing the binary tree. To achieve this goal, a chase decoding providing a list of candidate paths and satisfying node parity check formulas is used. ${\beta _j}$ denotes a candidate output codeword of target node. When starting from a source path $l$ with reliability $PM_l^{t - 1}$, the reliability of the path corresponding to the output codeword ${\beta _j}$ is
\begin{equation}
PM_{l,j}^t = PM_l^{t - 1} + \sum\limits_{i = 0}^{{N_v} - 1} {\left| {{\beta_j^t}[i] - h({\alpha _v^t}[i])} \right|\left| {{\alpha _v^t}[i]} \right|}
\label{equation05}
\end{equation}
Where $h()$ denotes hard-decision, $t$ denotes decoding step index, $N_v$ denotes the node size and ${\alpha _v}[i]$ is the input LLR values for the node.

\section{a simplified SCL polar decoder}
\subsection{Path metric probability table}
%

For SSCL decoder, the path metrics (PMs) in $t-1$ step are sorted in accending order. 
If the next node for decoding contains s information bits, there will be ${2^s}$ branches split from its parent path. We put PMs of these split branches into a {\bf{PM}} table row by row. Then, sort the elements in each row in ascending order and obtain matrix {\bf{PM'}}. $PM_{l,j}^t$ denotes the elements in l-th row and j-th column for t-th node. $PM_l^{t - 1}$ denotes the l-th PM obtained from (t-1) step.
Derived from (\ref{equation05}) that the elements in {\bf{PM'}} has the following propertites:
\begin{equation}
\begin{array}{c}
PM_{l,j}^t  =PM_l^{t - 1} + \sum\limits_i {\left| {{\beta _j}[i] - h({\alpha _v}[i])} \right|\left| {{\alpha _v}[i]} \right|}\\
PM_l^{t - 1} < PM_{l+1}^{t - 1}
\end{array}
\label{equation06}
\end{equation}
\begin{equation}
P{M_{l,j}^t} < P{M_{l,j + 1}^t}
\label{equation07}
\end{equation}
$l = 0,1,2,...,L - 1$, $j = 0,1,2,...$  

It should be noted that the {\bf{PM'}} is only used for illustrate the proposed method, and in next subsection we will show that the sorting operation to get {\bf{PM'}} is unnecessary for real implementation.

Although we don’t know the exact relationship of size between elements in {\bf{PM'}}. By (\ref{equation06}) and (\ref{equation07}), it can be found that the elements in the top left corner have a larger probability lower than the elements in the lower right corner. To verify that, numerical simulations are done to check the probability that each path is selected by a normal SCL decoder. These probabilities are written into a table called PM probability (PMP) table.

\begin{figure}[htb]
\centering
\includegraphics[scale=0.45]{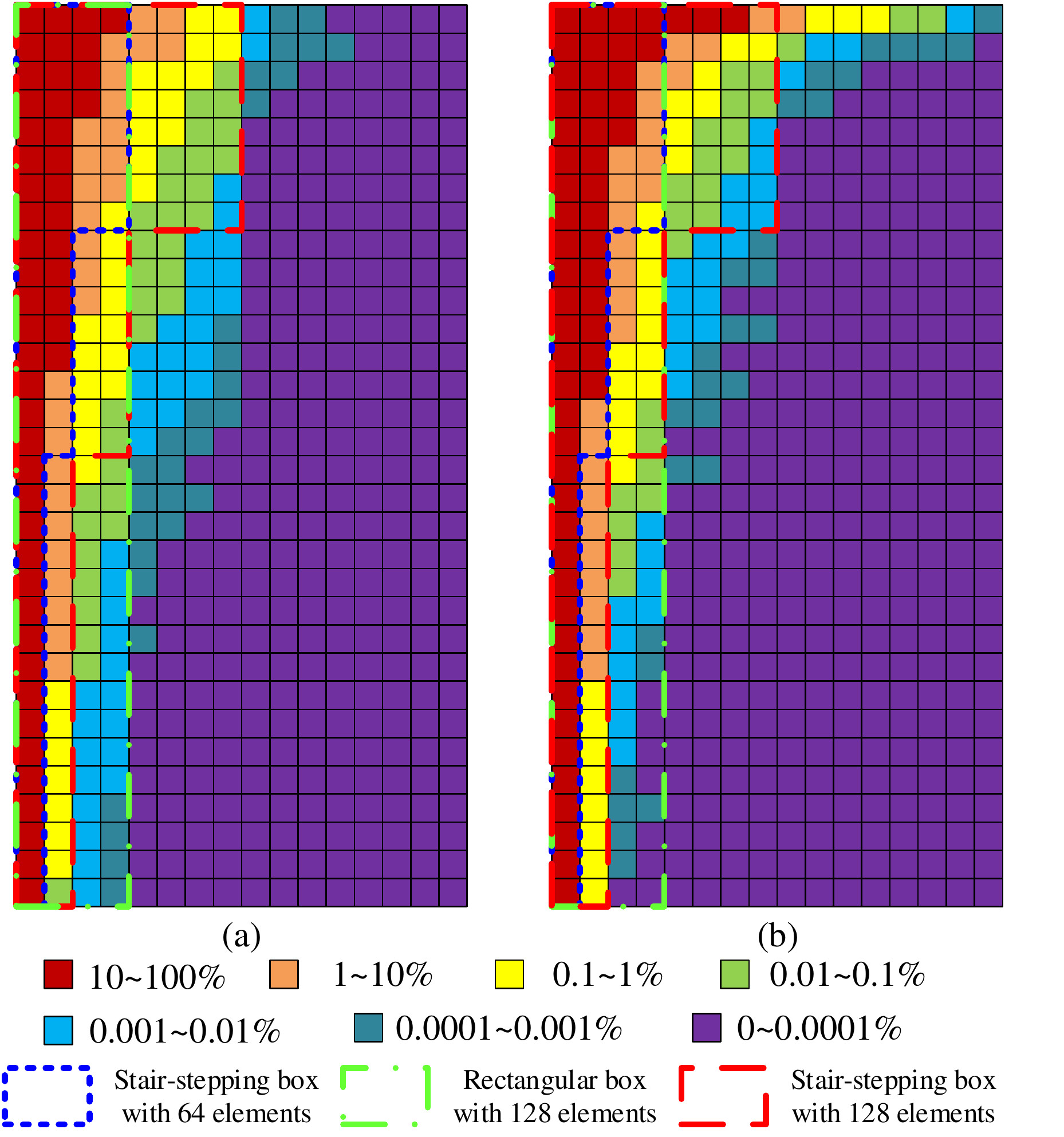}
\caption{
PM Probability table for all types of nodes together, $N=64$, $K_{payload}=32$, $K_{CRC}=11$ and $EsN0=3dB$.
(a) Node size $N_v$ is fixed to 4. 
(b) Node size $N_v$ is fixed to 8.
}
\label{AllNodePMPTable}
\end{figure}

Fig. \ref{AllNodePMPTable} shows a PMP table for $L=32$ as an example. The node size $N_v$ is fixed to 4 in Fig. \ref{AllNodePMPTable}(a) and 8 in Fig. \ref{AllNodePMPTable}(b). The probabilities of different type of nodes are drawn together. The unclassed nodes (not belong to Rate-1, Rate-0 and SPC) will be decoded by an ML decoder. Different colors represent different probabilities.The results show that SCL decoder is more likely to select the PMs in the top left corner on the ordered PM table, so a sorter for a part of PMs has potential to achieve comparable performance with a sorter for all PMs. 

Three boxes in different colors are used to select part of elements in the PMP table, and the sums of the probabilities are listed in Table \ref{table01}. The number (x, y, z) after the blue and red boxes represent that the boxes have three levels, each of which keeps x, y and z PMs. The ratio between three levels is set as 1:1:2. From the table, it is found that stair-stepping box is more efficient than a rectangular box for selecting the correct PMs.
\begin{table}[!t]
\renewcommand{\arraystretch}{1.3}
\caption{Probability for Different Path Selection Methods}
\label{table01}
\centering
\begin{tabular}{|c|c|c|c|c|}
\hline
\multirow{2}{*}{} &	\multirow{2}{0.6cm}{Node size} & \multicolumn{3}{c|}{Probability for paths selected by SCL decoder} \\
\cline{3-5}
  & &Blue box(4,2,1)&Green box(4)&Red box(8,4,2)\\
\hline
Fig.\ref{AllNodePMPTable}(a) & 4 & 98.26332\% & 99.33933\% & 99.97014\% \\
\hline
Fig.\ref{AllNodePMPTable}(b) & 8 & 96.44439\% & 97.64262\% & 99.86676\% \\
\hline
\end{tabular}
\end{table}

\begin{figure*}[!t]
\centering
\includegraphics[scale=0.75]{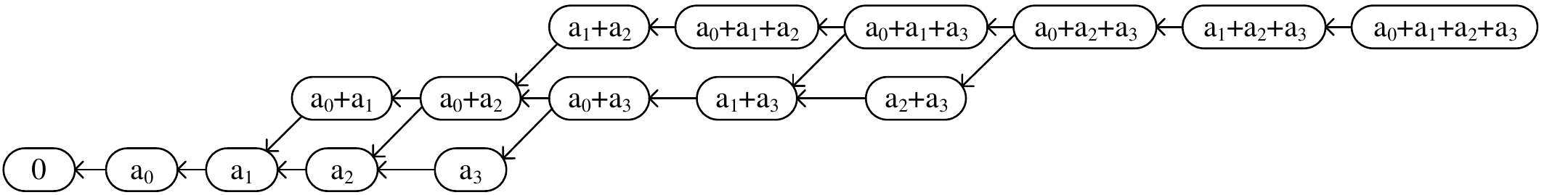}
\caption{
Directed graph for Rate1 node.
}
\label{4ParallelRate1PMRelations}
\end{figure*}



\subsection{A fast calculation of the PM table}

In Fig.\ref{AllNodePMPTable}, it is found that only the first few columns are needed for PS-SCL. In this part, we will illustrate how to get  the smallest 1, 2, 4 or 8 path metrics without sorting all the elements.

As (\ref{equation05}) shows, the updated PM value for the $t$ step can be calculated as a summation of the PM value in $t-1$ step and the BM (Branch Metric) value in . The BM is calculated as
\begin{equation}
BM_{l,j}^t = \sum\limits_{i = 0}^{{N_v} - 1} {\left| {{\beta _{l,j}^t}[i] - h({\alpha _l^t}[i])} \right|\left| {{\alpha _l^t}[i]} \right|}
\label{equation14}
\end{equation}
In which, $t$ is the operation step, $l$ is the path index, $N_v$ is the size of the node and $i$ is the bit index. ${\alpha _l}[i]$ is the input LLR of the i-th bit in subpath $l$. ${\beta _{l,j}}[i]$ donotes the i-th bit in sequence ${\beta _{l,j}}$, and ${\beta _{l,j}}$ denotes the $j$-th candidate output codeword for $l$-th path. $h(.)$ denotes hard-decision operation.

In the following text, we take ${N_v} = {L_s} = 4$ as an example to describe the BM selection procedure. The received LLR vector of a node is $\alpha [i](i = 0,1,2,3)$, and its absolute value vector is 
\begin{equation}
{a_i} = \left| {\alpha [\pi (i)]} \right|(i = 0,1,2,3)
\label{equation15}
\end{equation}
where $\pi \left( i \right)(i = 0,1,2,3)$ represents a permutation for $\left| {\alpha [i]} \right|$ in ascending order, which satisfies: 
\begin{equation}
0 \le {a_0} \le {a_1} \le {a_2} \le {a_3}
\label{relationsForRate1}
\end{equation}

For all the combination of the information bits, the BMs are calculated as:
\begin{equation}
B{M_j} = \sum\limits_{i = 0}^3 {\left| {{\beta _j}[\pi (i)] - h(\alpha [\pi (i)])} \right|{a_i}}
\label{equation16}
\end{equation}

If the information bit number ${N_i} = 4$ (Rate-1 node), all candidate results for BMs are listed as follows:
\begin{equation}
\begin{array}{*{20}{c}}
{0,}&{{a_2},}&{{a_3},}&{{a_0} + {a_1} + {a_3},}\\
{{a_0},}&{{a_0} + {a_2},}&{{a_0} + {a_3},}&{{a_0} + {a_2} + {a_3},}\\
{{a_1},}&{{a_1} + {a_2},}&{{a_1} + {a_3},}&{{a_1} + {a_2} + {a_3},}\\
{{a_0} + {a_1},}&{{a_0} + {a_1} + {a_2},}&{{a_2} + {a_3},}&{{a_0} + {a_1} + {a_2} + {a_3}}
\label{elementsInRate1Node}
\end{array}
\end{equation}

Here, a directed graph in Fig.\ref{4ParallelRate1PMRelations} is used to indicate the relationships between the PMs, in which an arrow pointing from A to B indicates that A is larger than B. Starting from any element in the graph, all other elements that can be reached are smaller than the element itself. So, we can find the smallest 8, 4, 2 or 1 elements by traversing the directed graph. For ${N_i} = 3$ (SPC node), similar graph can be used to find the specific  smallest elements. The results are shown in table \ref{table03}. For ${N_i} < 3$, there are only 4 or 2 candidate BMs, so 4-sorter or 2-sorter can be used directly to find the smallest 1 or 2 BMs.

\begin{table*}[!t]
\renewcommand{\arraystretch}{1.3}
\caption{Smallest Elements of Different Node Type}
\label{table03}
\centering
\begin{tabular}{|C{0.2cm}|C{2.3cm}|C{5cm}|C{2.5cm}|C{1.5cm}|C{1cm}|}
\hline
${N_i}$ & Syndrome & smallest 8 & smallest 4 & smallest 2 & smallest 1 \\ 
\hline
4 & - & ${0}$, ${a_0}$, ${a_1}$, ${a_0\!+\!a_1}$, ${a_2}$, ${a_0}\!+\!{a_2}$, ${min({a_3},{a_0}\!+\!{a_3},{a_1}\!+\!{a_2},{a_0}\!+\!{a_1}\!+\!{a_2})}$, ${submin({a_3},{a_0}\!+\!{a_3},{a_1}\!+\!{a_2},{a_0}\!+\!{a_1}\!+\!{a_2})}$ & ${0}$, ${a_0}$, ${a_1}$, $min({a_0\!+\!a_1},{a_2})$ & ${0}$, ${a_0}$ & ${0}$\\
\hline
\multirow{2}{*}{3} & $\sum\limits_{i = 0}^3 {h(\alpha [\pi (i)])}  = 0$ & - & ${0}$, ${a_0}\!+\!{a_1}$, ${a_0}\!+\!{a_2}$, $min({a_0}\!+\!{a_3},{a_1}\!+\!{a_2})$ & ${0}$, ${a_0}\!+\!{a_1}$ & ${0}$ \\
\cline{2-6}
 & $\sum\limits_{i = 0}^3 {h(\alpha [\pi (i)])}  = 1$ & - & ${a_0}$, ${a_1}$, ${a_2}$, $min({a_3},{a_0}\!+\!{a_1}\!+\!{a_2})$ & ${a_0}$, ${a_1}$ & ${a_0}$ \\
\hline
\end{tabular}
\end{table*}


\begin{figure}[!t]
\centering
\includegraphics[scale=0.5]{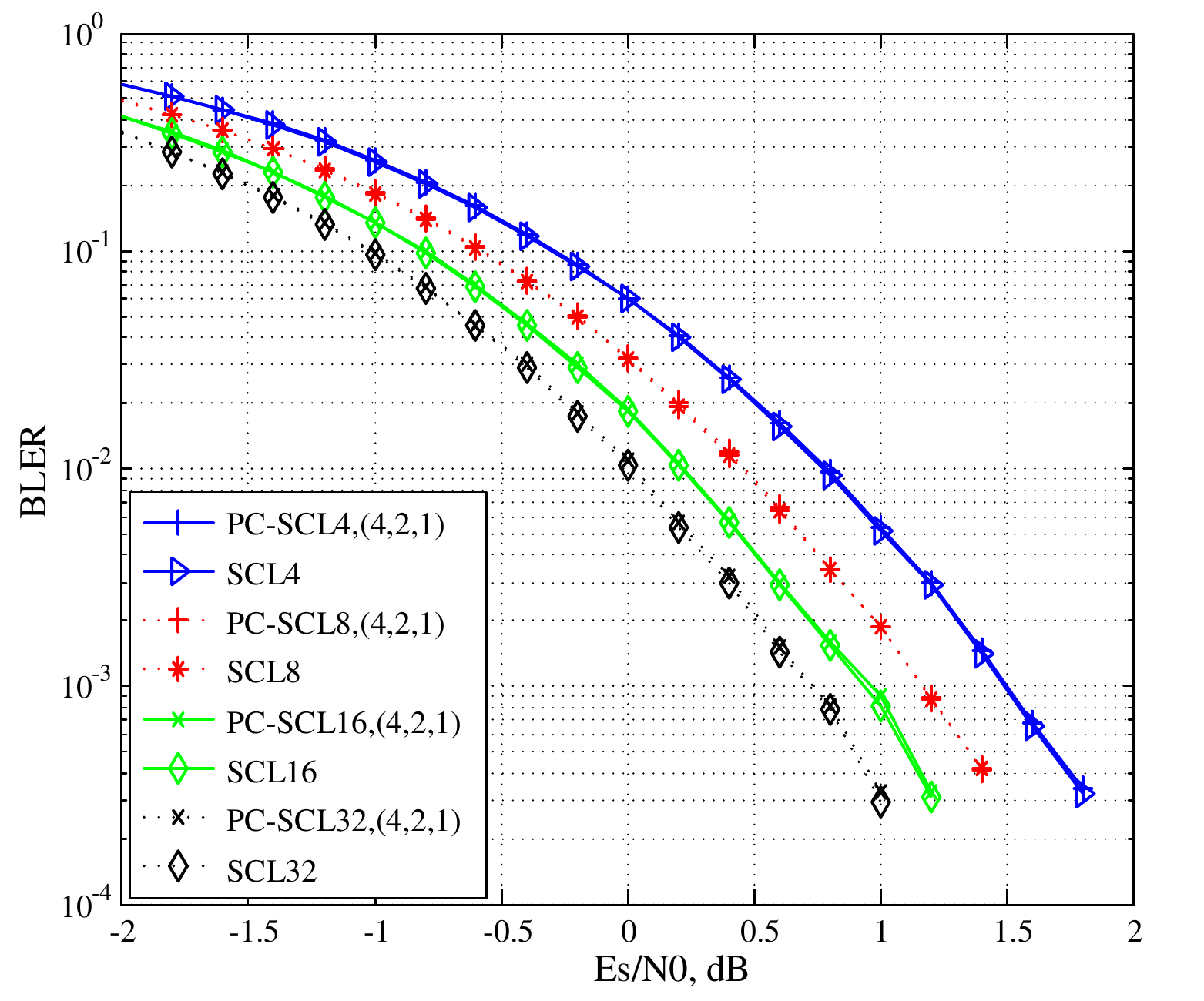}
\caption{
BLER for $N=128$, $K_{payload}=38$, $K_{CRC}=11$, $N_v=4$
}
\label{SimResultForParallel4}
\end{figure}

\begin{figure}[!t]
\centering
\includegraphics[scale=0.5]{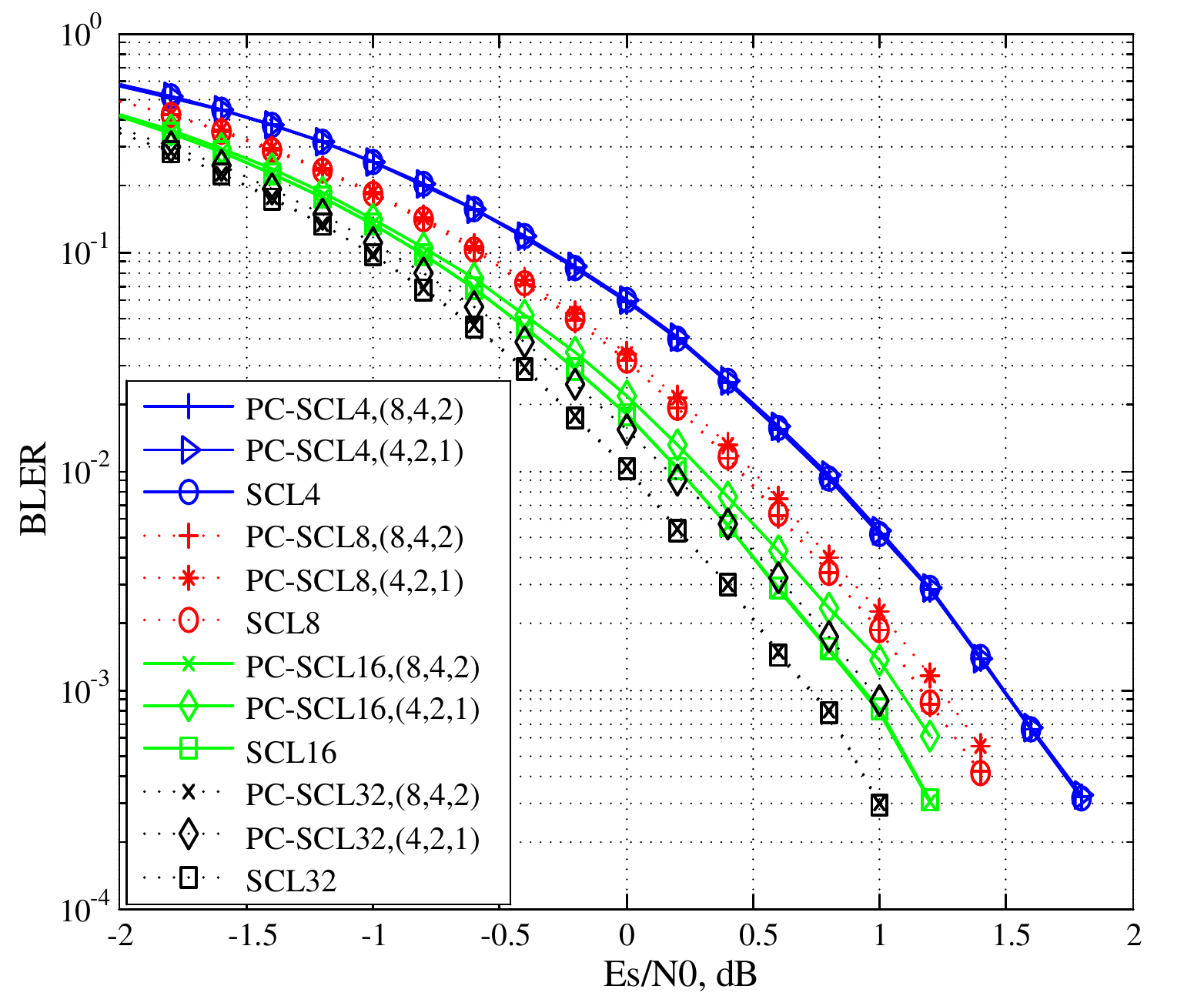}
\caption{
BLER for $N=128$, $K_{payload}=38$, $K_{CRC}=11$, $N_v=8$.
}
\label{SimResultForParallel8}
\end{figure}

\section{Numerical results}

Numerical simulation was conducted to test the error correction performance of the PS-SCL decoder. All simulation results are for QPSK modulated random codewords transmitted over an AWGN channel. The polar codes are constructed by PW sequence in \cite{bibitem11} with 38 bit payload, 11 bit CRC and are coded with 128 bits. In Fig.\ref{SimResultForParallel4}, the node size $N_v$ is fixed to 4. The PS-SCL method uses a stair-stepping selections with 3 different levels, and the number of kept PMs for the 3 levels is written as level (x, y, z). The ratio of the three levels is set as 1:1:2. For example, if PS-SCL32 is used with (4, 2, 1),  the kept PMs in ordered PM table should be the first 4 PMs from row 0 to row 7, the first 2 PMs from row 8 to row 15 and the first 1 PM from row 16 to row 31. The results show that at least for BLER up to 1e-4, the loss of PS-SCL decoder with level (4, 2, 1) compared with normal SCL decoder is negligible.

In Fig.\ref{SimResultForParallel8}, the node size $N_v$ is fixed to 8. For the PS-SCL decoder with level (4, 2, 1), as the list size changes from 4 to 32, its performance loss grows from 0.02dB to 0.3dB. If we increase the selection level to (8, 4, 2), the perfromance loss turns to be negligible. Because in Fig. \ref{AllNodePMPTable}, the probability distribution becomes more dispersed when node size grows, more PMs are needed to be selected to avoid performance loss.

\begin{table}[!t]
\renewcommand{\arraystretch}{1.3}
\caption{Number of CAS Units for Different Decoding Algorithm}
\label{table02}
\centering
\begin{tabular}{|C{0.8cm}|C{0.8cm}|C{0.8cm}|C{0.8cm}|C{0.8cm}|C{0.8cm}|C{0.8cm}|}
\hline
List Size &PS-SCL, $N_v = 4$, (4, 2, 1)&PS-SCL, $N_v = 8$, (8, 4, 2)& SCL, $N_v = 4$ & SCL, $N_v = 8$ &Fast-SSCL in \cite{bibitem10}, $N_v = 4$&Fast-SSCL in \cite{bibitem10}, $N_v = 8$ \\
\hline
L = 4& 24 & 80 & 672  & 28160 & 80 &80\\
\hline
L = 8& 80 & 240 & 1792  & 67584 & 240 &672\\
\hline
L = 16& 240 & 672 & 4608  & 159744 & 672 &1792\\
\hline
L = 32& 672 & 1792  & 11520  & 372736 & 1792 &4608\\
\hline
\end{tabular}
\end{table}

If a bitonic sorter is chosen, for $2{L_{in}}$  input PMs, the total number of compare-and-select (CAS) units can be calculated by (\ref{equation13}) from \cite{bibitem12}. The total numbers of CAS units for some SCL decoders are shown in Table \ref{table02}.
\begin{equation}
c_{tot}^{BT} = \frac{{{L_{in}}}}{2}({\log _2}{L_{in}} + 1)({\log _2}{L_{in}} + 2)
\label{equation13}
\end{equation}
Comparied with normal SCL, the proposed PS-SCL has 95\% less CAS units for node size 4 and 99\% less CAS units for node size 8. Comparied with Fast-SSCL, PS-SCL has 65\% less CAS units on average.

\section{Conclusion}
In this work, a selection method is proposed to reduce the complexity of sorting network for polar SCL decoder. The method takes advantage of the ordered property of path metrics (PMs) to make an ordered PM table, and uses a stair-stepping box selection on PM table to find the PMs that are more likely to be selected by a normal SCL decoder. Furthermore, a fast calculation method is proposed to find these PMs easily. Compared with normal SCL decoder, the PS-SCL can reduce the compare-and-select (CAS) units of a bitonic sorter up to 95\%. Numerical results show that the performance loss of the PS-SCL is negligible with a suitable selection box size. When list size or node size grows larger, a larger selection box size is needed to avoid error correction perfromance loss. Next, we will focus on designing different selecting boxes for different nodes in the future.


%





\ifCLASSOPTIONcaptionsoff
  \newpage
\fi

\end{document}